\documentclass{an}
\usepackage{graphicx}
\usepackage{times}
\usepackage{fancyhdr}
\begin{document}

\title{High frequency QPOs: nonlinear oscillations in strong gravity}

\author{W{\l}odek Klu\'zniak}
\institute{Zielona G\'ora University, ul Lubuska 2, 65-265 Zielona G\'ora,
Poland and\\
Copernicus Astronomical Center, Bartycka 18, 00-716 Warszawa, Poland}

\date{Received; accepted; published online}

\abstract{Some of the more promising ideas about the
origin of the high frequency variability (kHz QPOs) in the observed X-ray
emissions of low-mass X-ray binaries are contrasted with less promising ones. 
}

\correspondence{wlodek@camk.edu.pl}

\maketitle

\section{Setting the stage}
\subsection{Remembering the key crossroads}

If I am personally convinced that the high(est)-frequency QPOs in black hole
and neutron-star sources have a common origin, it is not only because
this assumption has served us well in the past five years, but also
because I have taken the other road, and I have seen that it leads nowhere.
I would like to begin by describing my journey, so that others may
decide whether or not the direction we now take is reasonable.

\subsection{The accretion-gap paradigm for neutron stars}

The discovery twenty years ago of the first millisecond (radio) pulsar
immediately brought home the idea that low-mass X-ray binaries (LMXBs)
may contain rapidly rotating neutron stars endowed with a magnetic field
far below the theretofore canonical value of $10^{12}\,$G.
These were hard-to-imagine times when the timing capability of
the the best X-ray instrument (GINGA) was below  $\sim200\,$ Hz,
and (EXOSAT) spectra were only reported up to 30 keV.

In our first paper on LMXBs, formally devoted to investigating
spin-up by accretion, we examined the general-relativistic
orbits around neutron stars with dynamically unimportant magnetic fields.
We have pointed out that for many viable equations of state (of matter
at supranuclear densities) the accretion disk could terminate close to the
marginally  stable orbit (ISCO) well outside the stellar surface
(Klu\'zniak and Wagoner 1985) and  dubbed this situation the
``accretion-gap regime.''
In other words, the accretion disk of a neutron star in a typical LMXB 
could well be very similar to the accretion disk around a black hole,
with the inner edge determined by GR effects of strong gravity (with
frame dragging and all).

The next paper in the short series, prepared as part of my thesis
work (Klu\'zniak 1987), demonstrated that the spectrum of the emission
expected in the accretion-gap regime,
when the freely falling matter which left the inner edge of the disk
hits the equatorial accretion belt, is a rather hard power-law extending
up to $\sim10^2\,$keV (Klu\'zniak and Wilson 1991).
The discovery several years later of such emissions
from some neutron star sources can be taken as possibly confirming
the presence of the ISCO.
However, non-thermal spectra do not uniquely point to the geometry or
process of emission, and so we also searched for timing signatures of
the accretion-gap regime or, as it is put today, of 
a key prediction of Einstein's theory of gravitation (GR).

We were looking not for a tiny quantitative departure from Newtonian
gravity, but for a major difference. The time of flight
of matter through the gap would be one such signature of a qualitative 
departure from Newtonian accretion geometry. Accordingly, we computed
the trajectory and the fate of a clump of matter orbiting
in the accretion disk, crossing the sonic point within the inner edge,
and hitting the stellar surface (Klu\'zniak, Michelson and Wagoner 1990).
We proposed this as a method of measuring the mass of the neutron star,
or (if the mass were known) of testing GR. In particular, interpreting 
the 0.2 Hz QPO observed in a certain X-ray pulsar
(Angelini, Stella and Parmar 1989) as (an unknown process)
occurring at the orbital frequency close to the inner boundary of the
accretion disk, which is $\sim10^3$ stellar radii above the surface of this
strongly magnetized accreting neutron star, we pointed out
that in a LMXB neutron star in the accretion gap regime the same
QPO should occur at a frequency which  a) is in the kHz range, b)
inversely depends on the mass $M$ of the neutron star, and c) depends
on the angular momentum $j$ of the neutron star---the frequency at
the ISCO being $2.2\, {\rm kHz}\, (M_\odot/M)(1+0.749 j)$.
It is in the context of the accretion-gap paradigm that our efforts
to understand kHz QPOs should be understood.

\subsection{Two kHz QPOs in neutron stars, one in black holes?}
The discovery in neutron stars of two kHz quasi-periodicities
in observations of neutron stars by the first instrument (RXTE)
that had the capability of observing such rapid variability
did not seem very surprising in light of
the Klu\'zniak, Michelson and Wagoner 1990 prediction discussed above.
Neither did the lack of a coherent pulsation associated with the spin
of the neutron star---after all, to obtain high orbital frequencies,
we had to assume that the stellar
magnetic field was too weak to channel the flow.
The frequencies came out just right, if the neutron stars were a bit
on the heavy side of the allowed mass range in an LMXB transferring
mass for $\sim10^8\,$y.

Soon a high frequency QPO was reported also from a black hole source.
This all seemed to make perfect sense. Black hole acretion disks were
just like neutron star acretion disks, the only difference being in
the size of their inner gap (proportional to mass). Hence, one QPO at
the inner edge of the disk was expected, probably associated with an
orbiting clump. On top of it, matter falling onto the equatorial
accretion belt differentially slipping on the neutron star
surface---i.e., rotating faster than the neutron star, albeit at
variable speeds---could support a structure (perhaps a vortex like the
great red spot on Jupiter, although the referee prevented this
specific suggestion from appearing in print) giving rise to the second
quasi-period. No such second QPO was expected from a black hole, where
matter fell down the horizon. At any rate, identifying the observed
frequency with that in the ISCO one could determine the mass of the
neutron star or black hole (Klu\'zniak 1998),
a favorite pastime for many authors in the subsequent years.

A model very similar to this (minus the strong gravity part),
 had been proposed much
earlier by Paczy\'nski to explain the white dwarf QPOs, and their
variable frequencies, so it all made perfect sense in the kHz QPO
context, where both frequencies were known to increase with the luminosity
of the source---the idea in the model being that at higher accretion rates
the sonic point moves in closer to the stellar surface, and that
the top of the equatorial accretion belt moves faster as the accretion rate
of angular momentum increases
(see  Klu\'zniak, Michelson and Wagoner 1990 for references
on the position of the sonic point).

However, when I tried to make the model of the frequency variation in kHz  QPOs
agree quantitatively with the data, I failed. And I could make
no sense of any other published model
based on the assumption that the presence of the second kHz QPO is a 
direct signature of the (spinning) stellar surface. A timely warning from
Dimitrios Psaltis that the final word on the number of QPOs in black holes
has not yet been said, as they are very weak at high frequencies, has
helped me to finally abandon the idea that only one of the high frequency QPOs
arises in the accretion disk. If one arises in the accretion disk,
why not both?

\subsection{Wrong turn}
Much initial work on netron-star QPOs was informed by earlier models
of similar phenomena in white dwarf systems (CVs), but the assumption
that the kHz QPOs were caused by orbiting clumps (inhomogeneities in
the accretion flow) turned out to be wrong.  It had been known that
clumps in the differentially rotating accretion disk would be sheared
out into nearly axisymmetric structures within a few orbits
(Bath, Evans and Papaloizou 1974)---i.e., a clump is short lived---and
 it has simply been overlooked
that the kHz QPOs were reported to be very
narrow features already in the discovery papers.
Today we know that the high quality factor,
$Q\sim200$ of the lower kHz QPOs in neutron-stars excludes all
clump-based models, as well as those involving longer-lived vortices
drifting radially inwards with the mean accretion flow (Barret et
al. 2005). Thus, the question of relative merits of the ``beat
frequency model'' and the ``relativistic precession model'' is a moot
point. 

\subsection{Linear diskoseismology}
All the while, fundamental work was being done on the normal modes
of accretion disks. It turns out, that in GR $g$-modes are trapped
near the maximum of the radial epicyclic frequency 
(Okazaki, Kato, Fukue 1987), an effect  clearly
absent in Newtonian $1/r$ gravity that requires the epicyclic frequencies
to be equal to the Keplerian orbital value $\sqrt{GM/r^3}$, itself
a monotonic function of the radius.
After nearly two decades of concerted effort (Wagoner 1999, Kato 2000)
 several modes have been identified, which were promising in the context
of black hole HFPOs.
Because this is such a difficult problem all work had been carried out in
the linear regime. 

One conclusion, more clearly seen {\sl a posteriori},
is that no special ratio of frequencies for any of the preferred low-lying
modes is predicted. More precisely, over and above the $1/M$ scaling
in GR, the frequencies of the various modes
are a function of the spin of the central compact object,
and so is the ratio of frequencies, at least for modes which are not harmonics
of the same fundamental. Thus, judiciously choosing the black hole spin
one can accommodate nearly any frequency ratio
of, e.g., the fundamental $g$-mode and $c$-mode. However, no two black holes
(in a limited sample) are expected to have the same spin, hence any such ratio
must be accidental and limited to one or two objects at most.

Another conclusion was that the frequencies have only a slight dependence
on the expected variations in the disk structure. Therefore, disk modes
were not thought to be directly responsible for modifying the X-ray flux
in neutron-star systems, where the QPO frequencies vary quite a bit in any
given source. 

These two conclusions need no longer hold in non-linear diskoseismology.
The equations of hydrodynamics are intrinsically non-linear and there is
urgent need to carry out an analysis of disk modes in the non-linear regime.
We have reasons to believe that the properties of the high frequency
QPOs in both black hole and neutron-star
systems are a manifestation of non-linear coupling between certain oscillations
of the accretion disk.

\section{Two characteristic variable frequencies}
The two kHz QPOs in neutron stars have reproducible properties in
each source, including characteristic frequency values, and yet the
two frequencies are not fixed---they vary. 
It is easy enough for the accreting fluid to display characteristic
frequencies. Even in Newtonian gravity, where the orbital frequency
is scale-free, there is a maximum frequency corresponding to that
in the inner disk---simply because there is a characteristic radius,
for example the stellar radius. In GR orbital motion characteristic
 frequencies are built in, for example the maximum radial epicyclic
frequency, and the frequencies of the trapped modes are a reflection
of this property of strong gravity. However, all these characteristic
frequencies are fixed for a given space-time metric (i.e., a given
neutron star or black hole).

It is the variation of frequencies that is
highly suggestive of non-linear dynamics. Frequencies that vary in a
characteristic range are a well studied feature of non-linear oscillators.
For these reason, we have suggested that the two kHz QPOs in neutron
stars are a manifestation of a nonlinear resonance in an accretion disk
placed in strong gravity (Klu\'zniak and Abramowicz 2001a,b). 
Two corrolaries immediately follow, and within a few months
both were directly verified by observers unaware of our hypothesis. 
The first is that there are preferred rational ratios of frequencies,
such as 1:2 or 2:3. The second is that high frequency QPOs in black holes
also come in pairs.

This triumph of simple physics invites more work to be done.
In paricular, it would be good to identify the two non-linear oscillators,
and it is necessary to convincingly identify the similarities and the
differences of black hole and neutron-star QPOs.
This is the subject of the current workshop.

\section{The smoking gun of non-linear oscillations}

The twin high-frequency QPOs need not have identically the same
propeties in black holes and neutron-stars.
 The differences,
if any, would presumably be linked to the presence of a rotating
magnetosphere in accreting neutron stars, which could directly imprint
the spin frequency on the oscillations of the disk. 

The stellar spin frequency $\nu_*$ may couple to another periodic motion
already present in the system in more than one way, easily leading
to the appearance of a beat frequency $\nu_2=\nu_1 \pm \nu_*$.
However, if  non-linearities are present, one-half of the spin frequency 
may appear in the data. We are fortunate in that nature has provided us
with transient systems where the accreting neutron star displays
a coherent frequency, surely equal to its spin frequency. In the first
of these systems to be closely studied, SAX J1808,
two kHz QPOs have been revealed,
whose frequencies satisfy $\nu_2-\nu_1=\nu_*/2$.
In our view, this is a clear signature of non-linearities in the system
(Klu\'zniak et al. 2004).

We may even try to give a specific model for the non-linear interaction.
Elsewhere, we have shown that certain fluid configurations in the accretion
flow, e.g. tori, can execute two quasi-rigid
oscillatory motions accurately described
by equations nearly identical to that of a test particle in a circular
orbit, with the mode frequencies equal to the epicyclic frequencies
in a certain orbit (Klu\'zniak and Abramowicz 2002, Lee et al. 2004).
Let us focus on one of these oscillations, with the radial epicyclic frequency
$\omega_r$. In the presence of the neutron star driver,
it could be described by the equation of a non-linear forced oscillator,
with a forcing function $f(t)$ that is periodic at the spin frequency,
$\nu_*$:
$$\ddot r +\omega_r^2 r +a(z)r^2=f(t).\eqno(1)$$
The configuration is also executing a vertical motion, whereby the value
of the vertical co-ordinate $z$ in this equation oscillates at the frequency
$\omega_z$. We identify the observed frequencies with the frequencies
of the oscillators: $\nu_1=\omega_r/(2\pi)$ and $\nu_2=\omega_z/(2\pi)$.
Reflection symmetry in the equatorial plane requires the expansion of the
$a(z)$ coefficient to only have even powers in $z$. The leading 
coupling term, then, has the form $a_2 z^2r^2$, with $a_2$a constant.
Substituting the harmonic time dependence of $z$, $r$, and $f$ at frequencies
respectively, $\nu_2$, $\nu_1$, and $\nu_*$, we see that the condition
for forced resonance is $2(\nu_2\pm\nu_1)=\nu_*$.
For realistic models of the spinning neutron star in J1808, the option
$\nu_2-\nu_1=\nu_*/2$ is viable. Some other possible forms of the non-linear
equation also yield the same result.

\section{A model for the 3:2 frequency ratio}

The same oscillation model, even without the forcing term and the $r^2$
non-linearity, allows a simple explanation for the 3:2 frequency ratio
that is observed in the high frequency QPOs in black holes, as was noted in
Abramowicz and Klu\'zniak 2001, Remillard et al. 2002, and as we argue,
is also present in neutron stars.
The epicyclic motions of the fluid configurations,
e.g., of a torus are given by the equations
$$\ddot r +\omega_r^2 r =0, \eqno(2)$$
$$\ddot z +\omega_z(r)^2 z =0. \eqno(3)$$
In view of reflection symmetry in the $z=0$ plane,
 we have neglected the $z$ variation of the radial epicyclic eigen-frequency.
If only one oscillation is present, the frequencies are taken at a fixed
equilibrium point, so that the solution is a simple harmonic oscillator
motion. Now imagine that both oscillations are present. The radial motion
induces harmonic variations of the eigenfrequency $\omega_z$,
occuring at the frequency $\omega_r$, 
and the vertical motions can be resonantly 
excited by the horizontal oscillations. This is a situation very familiar to
any child exciting a swing, formally the equation is known as the Mathieu
equation, and the resonance (known as the parametric resonance) occurs
when
$$\omega_r={2\over n}\omega_z, \eqno(4)$$
with $n=1,2,3...$ an integer.
In general relativity, the first and strongest resonance occurs for $n=3$,
leading to a 3:2 ratio of the eigenfrequencies
(Klu\'zniak and Abramowicz 2002), because, in additon to eq. (4), the relation
$$\omega_r<\omega_z$$ must be satisfied.

\section{The Bursa plot}

The final question that must be discussed is whether neutron-star QPOs
are indeed the same phenomenon as the black hole ones. This is not
immediately obvious because the kHz frequencies in neutron stars vary.
In fact the frequency ratio of these kHz QPOs departs from 3:2.
There are three points to be made. 

One is that the histogram
of frequency ratios for individual neutron star sources, as well as for
all of them jointly, peaks near the value 1.5
(Abramowicz, et al. 2003a).
This is apparent in the inset of the figure
(where actually the ratio of the lower frequency to the higher is plotted,
so that the peak is close to 0.67).
The black hole QPOs are much weaker so, at present, we only see the peak
of the distribution. It is not know whether in black holes
 the weaker QPOs, presumably also
present, have the same frequency, or whether their frequencies
also depart from the 3:2 line.

The second point is that the departure of the observed frequencies
from the eigenfrequencies of the system, and of the ratio from the
resonant 3:2 ratio, is a generic feature of non-linear resonance
in coupled oscillators, so the line of correlation for an individual
neutron star can be reproduced in numerical and analytic work
(Abramowicz et al. 2003b, Rebusco 2004, Horak 2004).

The third point, as is very apparent in the figure, is that the
data points for at least some neutron stars seem to lie along straight
lines on the frequency-frequency plot, with a common intersection
point. This intersection point lies on the 3:2 line and corresponds
to frequencies which are higher than the black hole frequencies.
This can be explained by assuming that these neutron stars have very
similar masses, smaller than the (more varied) masses of the black holes.
The latter already show a $1/M$
 frequency dependence discovered by Remillard and
McClintock. The common intersection point of (some of) the neutron
star lines greatly strengthens the evidence for the 3:2 resonance in
neutron star QPOs, as well as for the strong gravitational field origin
of the high frequency QPO phenomenon.

\begin{figure}
\resizebox{\hsize}{!}
{\includegraphics[]{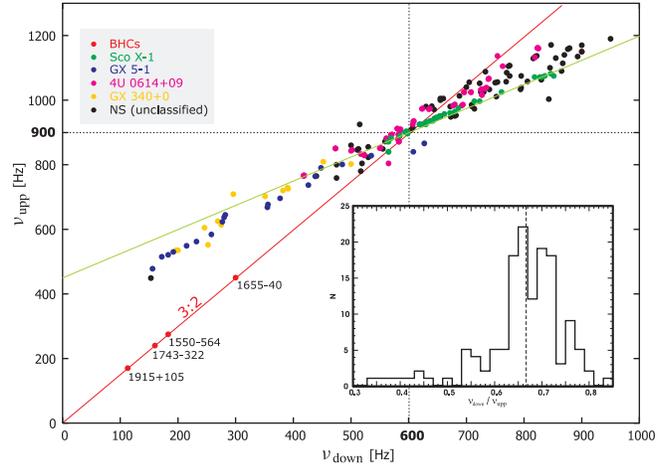}}
\caption{The Bursa plot. The highest QPO frequency is plotted against
its lower frequency counterpart. The black hole frequencies are from Remillard
and McClintock 2005. The neutron star twin kHz QPO frequencies are based on
published data (see Abramowicz et al. 2003a for references).
Note that the neutron star points are scattered about a line
intersecting the line of a 3:2 ratio in a single point $\sim$ (600 Hz, 900 Hz),
and that all black hole frequencies are lower, consistent with
 a $1/M$ dependence.
The green line going through the Sco X-1 data has a slope of 3/4.
 This suggests a simple
non-linear physical mechanism for its origin.
}
\label{figlabel}
\end{figure}

\acknowledgements
Supported in part by KBN grant 2P03D01424, and by the hospitality of NORDITA.

\end{document}